\documentstyle[12pt,epsfig]{article}
\begin{document}

\begin{center}
{\Large {\bf ZnWO$_4$ crystals as detectors for 2$\beta$ decay and
dark matter experiments}}
\end{center}

\vskip 0.2cm

\begin{center}

F.A.~Danevich\footnote{Corresponding author. {\it E-mail address:}
danevich@kinr.kiev.ua}, V.V.~Kobychev, S.S.~Nagorny, D.V.~Poda,
V.I.~Tretyak, S.S.~Yurchenko,
Yu.G.~Zdesenko\footnote{Deceased}

\noindent{\it Institute for Nuclear Research, MSP 03680 Kiev,
Ukraine}

\end{center}

\vskip 0.4cm

\begin{abstract}
\noindent
Energy resolution, $\alpha/\beta$ ratio, and the pulse shape
discrimination ability of the ZnWO$_4$ crystal scintillators were
studied. The radioactive contamination of a ZnWO$_4$ crystal was
investigated in the Solotvina Underground Laboratory.
Possibilities to apply ZnWO$_4$ crystals for the dark matter and
double beta decay searches are discussed. New improved half-life
limits on double beta decay in zinc isotopes were established, in
particular, for $\varepsilon\beta^+$ decay of $^{64}$Zn as:
$T_{1/2}^{2\nu}\geq 8.9\times 10^{18}$ yr and $T_{1/2}^{0\nu}\geq
3.6\times 10^{18}$ yr, both at 68\% CL.
\end{abstract}

\vskip 0.4cm


\vskip 0.4cm


\vskip 1.0cm

\section{INTRODUCTION}

Observation of the neutrinoless (0$\nu$) double beta ($2\beta$)
decay and/or dark matter particles would be extraordinary events
for the modern physics. It is because both the phenomena are
related to the new physical effects beyond the Standard Model of
the particle theory.

The great interest to the $0\nu2\beta$ decay searches
\cite{Ver02,Zde02,Ell02,DBD-table} have arisen from the recent
evidence of neutrino oscillations, strongly suggesting that
neutrinos have nonzero mass. While oscillation experiments are
sensitive to the neutrinos mass difference, only the measured
$0\nu2\beta$ decay rate could establish the Majorana nature of the
neutrino and give the absolute scale of the effective neutrino
mass.

Several sensitivity studies of the $2\beta$ decay processes were
performed using low background crystal scintillators, which
contain various candidate nuclei: $^{40}$Ca \cite{Bel99},
$^{48}$Ca \cite{Mat66,Oga04}, $^{106}$Cd \cite{Dan96,Dan03},
$^{108}$Cd, $^{114}$Cd and $^{116}$Cd \cite{Dan03}, $^{136}$Ce and
$^{138}$Ce \cite{Bel03}, $^{160}$Gd \cite{GSO}, $^{180}$W and
$^{186}$W \cite{Dan02,Dan03}. For instance, in the experiment
\cite{Dan03} with the cadmium tungstate (CdWO$_4$) scintillators
enriched in $^{116}$Cd the very low counting rate of 0.04
counts/(yr$\cdot$keV$\cdot$kg) was reached in the energy window
$2.5-3.2$ MeV. The half-life limit on $0\nu2\beta$ decay of
$^{116}$Cd was set as $T_{1/2}^{0\nu2\beta}\geq1.7\times10^{23}$
yr at 90\% CL, which leads to one of the strongest restriction on
the effective Majorana neutrino mass $\langle m_{\nu} \rangle\leq
1.7$ eV.

There are strong evidence for a large amount of dark matter in the
Universe (matter which reveals itself only through gravitational
interaction). Weakly interacting massive particles (WIMP) -- in
particular neutralino, predicted by the Minimal Supersymmetric
extensions of the Standard Model -- are considered as one of the
most acceptable component of the dark matter \cite{DM}. These
particles can be detected due to their scattering on nuclei
producing low energy nuclear recoils. Extremely low counting rate
and small energy of recoil nuclei are expected in experiments to
search for the WIMP. Therefore, the experiments for direct WIMP
search require extremely low level of background (less than
several events per kilogram per day per keV) with very low energy
threshold (less than tens keV).

Direct methods of WIMP detection are based on registration of
ionization or/and excitation of recoil nucleus in the material of
the detector itself. At present, most sensitive experiments apply
different detectors for WIMP search: Ge semiconductor detectors
\cite{Heidelberg,IGEX,Kli98}, bolometers
\cite{TeO2,LiF_bolometer}, scintillation detectors
\cite{DAMA,NaI-Spain,UKDMC,NaI-France,CaF2-DAMA,CaF2-Tokio,Xe-DAMA}.
An interesting possibility to reject background caused by
electrons provides cryogenic technique, which uses simultaneous
registration of heat and charge \cite{EDELWEISS,CDMS} or heat and
light signals \cite{Bobin97,CRESST,ROSEBUD}. In Ref.
\cite{CRESST,ROSEBUD} calcium tungstate crystals (CaWO$_4$) are
discussed as promising material for such a kind of dark matter
detector. However, radioactive contamination of CaWO$_4$ crystals
used in \cite{CRESST,ROSEBUD}, as well as investigated in
\cite{Zde04}, is too high for low counting
measurements.\footnote{See, however, \cite{CARVEL}.}

The purpose of our work was investigation of scintillation
properties and radioactive contamination of zinc tungstate
(ZnWO$_4$) crystals as possible detectors for the double beta
decay and dark matter experiments.

\section{MEASUREMENTS AND RESULTS}

\subsection{Light output and energy resolution of the ZnWO$_4$ scintillators}

Three clear, slightly colored ZnWO$_4$ crystals
($\oslash14\times4$ mm, $\oslash14\times7$ mm, and
$\oslash13\times28$ mm) were used in our studies. The crystals
were fabricated from monocrystals grown by the Czochralski method.
The main properties of ZnWO$_4$ scintillators are presented in
Table 1, where characteristics of calcium and cadmium tungstates
are given for comparison.

\begin{table}[htb]
\caption{Properties of ZnWO$_4$, CaWO$_4$ and CdWO$_4$ crystal
scintillators}
\begin{center}
\begin{tabular}{|l|l|l|l|}
 \hline
  ~                                     & ZnWO$_4$ & CaWO$_4$ & CdWO$_4$  \\
 \hline
 Density (g/cm$^3$)                     &  7.8          & 6.1           & 8.0    \\
 Melting point ($^\circ$C)              &  1200         & $1570-1650$   &  1325   \\
 Structural type                        &  Wolframite   & Sheelite      & Wolframite \\
 Cleavage plane                         & Marked (010)  & Weak (101)    & Marked (010) \\
 Hardness (Mohs)                        &  $4-4.5$      & $4.5-5$       & $4-4.5$     \\
 Wavelength of emission maximum (nm)    &      480      &   $420-440$   & 480     \\
 Refractive index                       &     $2.1-2.2$ & 1.94          & $2.2-2.3$  \\
 Effective average decay time$^{\ast}$ ($\mu$s) &   24  & 8             & 13  \\
 Photoelectron yield [\% of NaI(Tl)]$^{\ast}$  &  13\%     &      18\%  & 20\%    \\
 \hline
 \multicolumn{3}{l}{$^{\ast}$For $\gamma$ rays, at room temperature.} \\
\end{tabular}
\end{center}
\end{table}

Scintillation characteristics of zinc tungstate
crystals have been studied in \cite{Grab84,Zhu86,Holl88,Dan89}.
Light yield of ZnWO$_4$ was measured in \cite{Holl88} with the
help of silicon photodiodes as 9300 photons/MeV (which is
$\approx23$\% of NaI(Tl) light yield). We have estimated the value
of the photoelectron yield of the ZnWO$_4$ in measurements with
two scintillators: ZnWO$_4$ $\oslash14\times7$ mm and standard
NaI(Tl) $\oslash40\times40$ mm. The crystals were coupled to the
PMT (EMI D724KFLB) with the bialkali RbCs photocathode. The
ZnWO$_4$ detector was wrapped by PTFE reflector tape. The relative
photoelectron yield of ZnWO$_4$ was determined as 13\% of that of
NaI(Tl).

The energy resolution of ZnWO$_4$ crystals for 662 keV $\gamma$
rays ($^{137}$Cs) has been reported as FWHM$=13\%$ \cite{Zhu86}
and FWHM$=11.5\%$ \cite{Dan89}. In the present work the energy
resolution FWHM$~=11.0\%$ for 662 keV $\gamma$ line of $^{137}$Cs
was measured with the ZnWO$_4$ $\oslash14\times7$ mm crystal. The
crystal was wrapped by PTFE reflector tape and optically coupled
to 3" photomultiplier (PMT) Philips XP2412. Taking into account
slow decay of ZnWO$_4$ scintillation, the 24 $\mu$s shaping time
of spectroscopy amplifier was set.

A substantial improvement of light collection ($\approx40$\%) and
energy resolution was achieved by placing the crystal in liquid
(silicone oil with index of refraction $\approx1.5$). The crystal
was fixed in center of the teflon container $\oslash 90 \times 70$
mm and viewed by two PMTs XP2412. Fig. 1(a) demonstrates the
energy spectra of $^{137}$Cs and $^{60}$Co obtained in such a way
with the $\oslash14\times $7 mm ZnWO$_4$ scintillation crystal.
For the first time the energy resolution 9.1\% and 6\% was
measured with ZnWO$_4$ crystal scintillator for 662 and 1333 keV
$\gamma$ lines, respectively. Furthermore, the resolution 11.3\%
for the 662 keV $\gamma$ line was obtained with the larger crystal
$\oslash 13\times $28 mm in the same measurement conditions.

\nopagebreak
\begin{figure}[ht]
\begin{center}
\mbox{\epsfig{figure=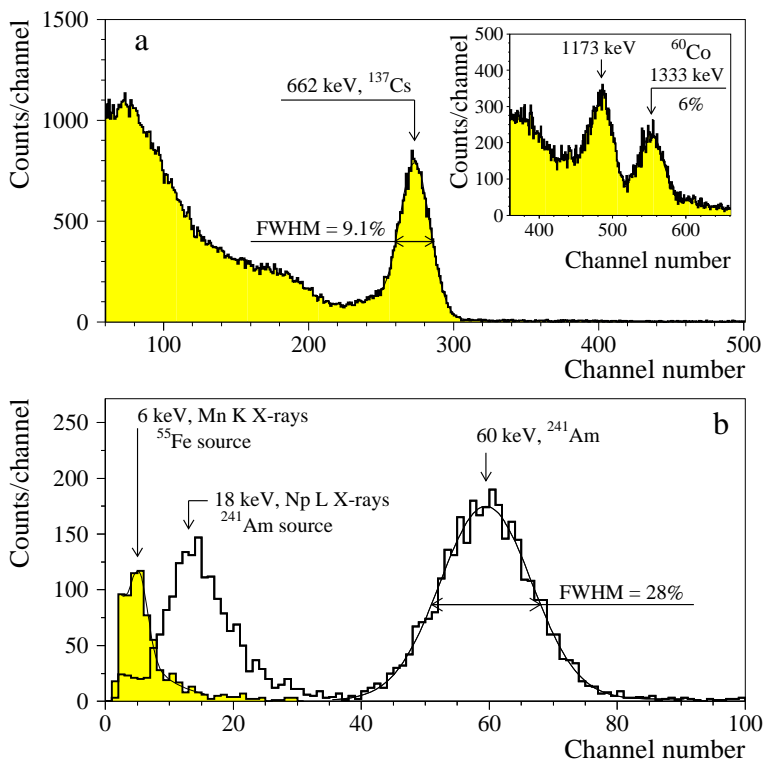,height=10.0cm}}
\caption {(a) Energy spectra of $^{137}$Cs and $^{60}$Co
(Inset) $\gamma$ rays measured with ZnWO$_{4}$ scintillation
crystal $\oslash 14\times7$ mm located in liquid and viewed by two
distant PMTs. (b) Energy spectra of $^{241}$Am and $^{55}$Fe low
energy gamma/X-rays measured by ZnWO$_{4}$ scintillation crystal
($\oslash 14\times 4$ mm).}
\end{center}
\end{figure}

Fig. 1(b) shows the energy spectra of $^{241}$Am and $^{55}$Fe low
energy gamma and X-ray lines measured with the ZnWO$_4$
scintillator $\oslash 14\times 4$ mm (the crystal was wrapped by
PTFE reflector tape and optically coupled to the PMT XP2412). The
6 keV peak of $^{55}$Fe source is still resolved from the PMT
noise.

\subsection{$\alpha /\beta$ ratio}

The $\alpha /\beta$ ratio\footnote{The $\alpha /\beta $ ratio is
defined as ratio of $\alpha $ peak position in the energy scale
measured with $\gamma $ sources to the energy of $\alpha $
particles. Because $\gamma$ quanta interact with detector by
$\beta$ particles, we use more convenient term ''$\alpha /\beta$
ratio''.} was measured (in the energy range of $1.0-5.3$ MeV) with
the help of the collimated beam of $\alpha$ particles from an
$^{241}$Am source and using various sets of thin mylar films as
absorbers. The energies of $\alpha$ particles were determined with
the help of a surface-barrier detector (see for details
\cite{W-alpha}). The ZnWO$_{4}$ crystal was irradiated in the
directions perpendicular to (010), (001), and (100) crystal
planes. The dependence of the $\alpha /\beta $ ratio on energy and
direction of $\alpha$ beam relatively to the crystal planes (see
Fig.~2) is similar to that observed with CdWO$_4$ scintillators
\cite{W-alpha}. As the quenching of the scintillation  light
caused by $\alpha$ particles (in comparison with electrons) is due
to the higher ionization density of $\alpha$ particles, such a
behaviour of the $\alpha /\beta $ ratio can be explained by the
energy dependence of ionization density of $\alpha$ particles
\cite{Birks}.

\nopagebreak
\begin{figure}[ht]
\begin{center}
\mbox{\epsfig{figure=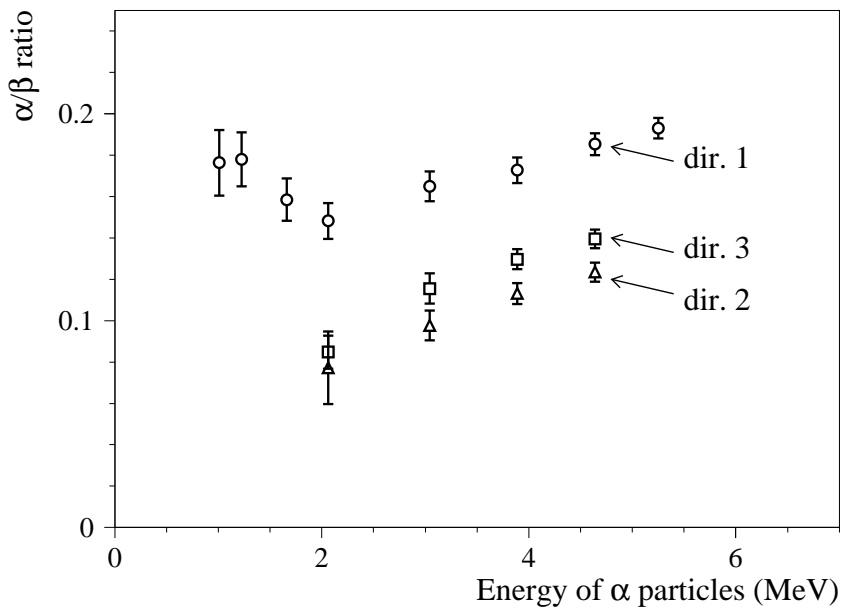,height=8.0cm}}
\caption {The dependence of the $\alpha /\beta $ ratio on
energy of $\alpha $ particles measured with ZnWO$_{4}$
scintillator. The ZnWO$_{4}$ crystal $\oslash 14\times 7$ m was
irradiated in the directions perpendicular to (010), (001) and
(100) crystal planes (directions 1, 2 and 3, respectively).}
\end{center}
\end{figure}

\subsection{Pulse shape discrimination}

The time characteristics of ZnWO$_4$ scintillators were studied as
described in \cite{Faz98,W-alpha} with the help of a transient
digitizer based on the 12 bit ADC (AD9022) operated at the sample
rate of 20~MS/s. The scintillator was irradiated by $\approx4.6$
MeV $\alpha$ particles in the direction perpendicular to the (010)
crystal plane. Shapes of the light pulses produced by $\alpha$
particles  and $\gamma$ rays are depicted in Fig.~3. Fit of the
pulses by the function: $f(t)=\sum A_{i}/(\tau_{i}-\tau_{0})\times
(e^{-t/\tau _{i}}-e^{-t/\tau _{0}}), ~ t>0$, where $A_{i}$ are
intensities (in percentage of the total pulse intensity),
$\tau_{i}$ are decay constants for different light emission
components, $\tau_{0}$ is integration constant of electronics
($\approx 0.2~\mu$s), gives three decay components
$\tau_{i}\approx0.7$, $\approx7$ and $\approx25$ $\mu$s with
different amplitudes for $\gamma$ rays and $\alpha$ particles (see
Table 2). The value of the slow decay component is in agreement
with the result of 25 $\mu$s obtained in \cite{Zhu86}, while 0.7
and 7~$\mu$s decay time constants were identified at the first
time. We were not able to measure the 0.1 $\mu$s decay component
reported in \cite{Zhu86}, because of the not enough fast digitizer
used in our measurements.

\nopagebreak
\begin{figure}[ht]
\begin{center}
\mbox{\epsfig{figure=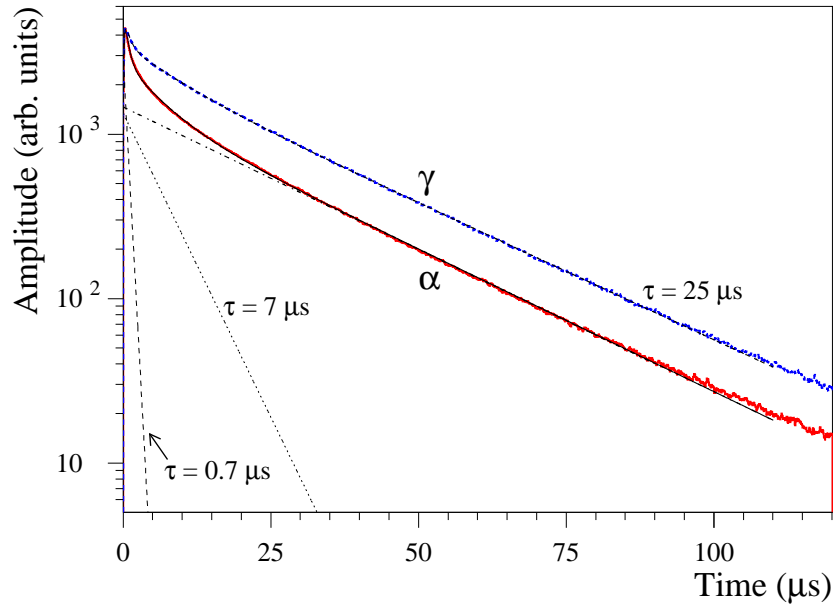,height=8.0cm}}
\caption {Decay of scintillation in ZnWO$_4$ for $\gamma$
rays and $\alpha$ particles (5 thousand forms of $\gamma$ and
$\alpha$ signals were added) and their fit by three components
with $\tau_i\approx0.7$, $\approx7$, and $\approx25$ $\mu$s.}
\end{center}
\end{figure}

\begin{table}[htb]
\caption{Decay times of ZnWO$_4$ scintillators for $\gamma$ quanta
and $\alpha$ particles. The decay constants and their intensities
(in percentage of the total intensity) are denoted as $\tau_i$ and
A$_i$, respectively.}
\begin{center}
\begin{tabular}{|l|l|l|l|l|}
\hline
 Type of irradiation  &  \multicolumn{3}{|c|}{Decay constants, $\mu$s}  \\
 \cline{2-4}
  ~   & $\tau_1$~(A$_1$)  & $\tau_2$~(A$_2$) & $\tau_3$~(A$_3$)\\
\hline
 $\gamma$ ray & 0.7~(2\%) & 7.5~(9\%) & 25.9~(89\%)    \\
 $\alpha$ particles & 0.7~(4\%) & 5.6~(16\%) & 24.8~(80\%)    \\
\hline
\end{tabular}
\end{center}
\end{table}

The difference of the pulse shapes allows to discriminate $\gamma
$($\beta$) events from those of $\alpha$ particles. For this
purpose we used the optimal filter method proposed in
\cite{Gatti}, which was successfully applied to CdWO$_4$
scintillators \cite{Faz98}. To obtain the numerical characteristic
of scintillation signal, so called shape indicator ($SI$), the
following procedure was applied for each pulse: $SI=\sum
f(t_k)\times P(t_k)/\sum f(t_k)$, where the sum is over time
channels $k,$ starting from the origin of pulse and up to 90 $\mu
$s, $f(t_k)$ is the digitized amplitude (at the time $t_k$) of a
given signal. The weight function $P(t)$ is defined as:
$P(t)=\{\overline{f}_\alpha (t)-\overline{f}_\gamma
(t)\}/\{\overline{f}_\alpha (t)+\overline{f}_\gamma (t)\}$, where
$\overline{f}_\alpha (t)$ and $\overline{f}_\gamma (t)$ are the
reference pulse shapes for $\alpha$ particles and $\gamma$ quanta.
Clear discrimination between $\alpha$ particles and $\gamma$ rays
($\beta$ particles) was achieved using this approach, as one can
see in Inset of Fig.~4 where the shape indicator distributions
measured by the ZnWO$_{4}$ scintillation crystal with $\alpha$
particles ($E_{\alpha}\approx 5.3$ MeV) and $\gamma$ quanta
($\approx 1$ MeV) are shown. As a measure of the discrimination
ability, the following relation can be used:
 $m_{PSA}=\mid SI_{\alpha}-SI_{\gamma}\mid/\sqrt{\sigma_{\alpha}^2+\sigma_{\gamma}^2}$,
where $SI_{\alpha}$ and $SI_{\gamma}$ are mean $SI$ values of
$\alpha$ particles and $\gamma$ quanta distributions (which are
well described by Gaussian functions), $\sigma_{\alpha}$ and
$\sigma_{\gamma}$ -- corresponding standard deviations. For the
distributions presented in Inset of Fig.~4, parameter
$m_{PSA}=5.2$.

\nopagebreak
\begin{figure}[ht]
\begin{center}
\mbox{\epsfig{figure=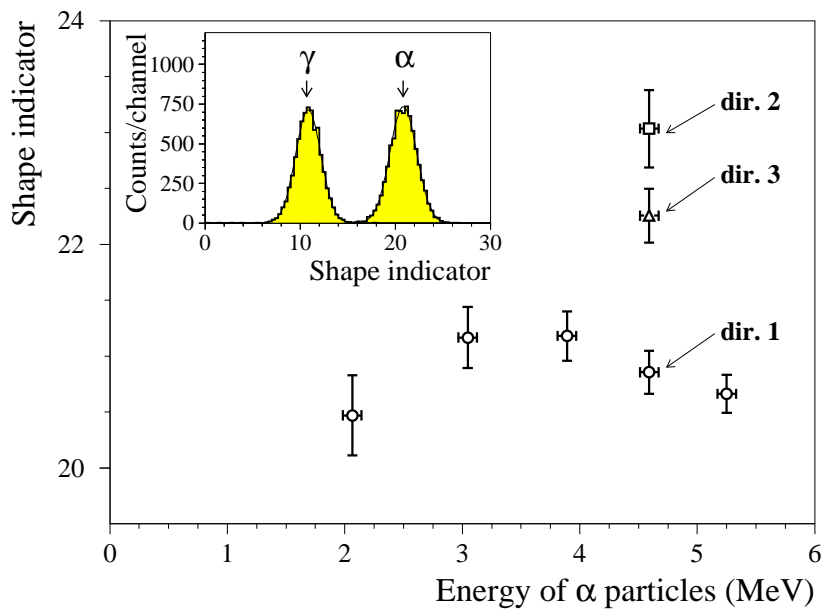,height=8.0cm}}
\caption {Dependence of the shape indicator (see text) on
the energy and directions of $\alpha$ particles relatively to the
main (010), (001) and (100) crystal planes of ZnWO$_{4}$
(directions 1, 2, and 3, respectively). (Inset) The shape
indicator distributions measured by ZnWO$_4$ detector with 5.3 MeV
$\alpha$ particles and $\approx$1 MeV $\gamma$ quanta.}
\end{center}
\end{figure}

The energy dependence of the $SI$ was studied with the $\oslash
14\times 7$ mm ZnWO$_4$ crystal in the $2-5.3$ MeV region for
alpha particles and $0.1-2.6$ MeV for gamma quanta. No dependence
of the $SI$ on the energy of $\gamma$ quanta was observed. The
measured dependence of the $SI$ on energy and direction of
irradiation by $\alpha$ particles (see Fig.~4) is similar to that
observed with CdWO$_4$ scintillators \cite{W-alpha}.

\subsection{Radioactive contamination}

Radioactive contamination of the ZnWO$_4$ crystal
($\oslash14\times4$ mm, mass of 4.5 g) was measured in a low
background set-up installed in the Solotvina Underground
Laboratory built in a salt mine 430 m underground ($\simeq$1000 m
of water equivalent) \cite{Zde87}. In the set-up the ZnWO$_4$
scintillation crystal  was viewed by the 3'' PMT (FEU-137) through
the high pure quartz light-guide 4.9 cm in diameter and 25 cm
long. The detector was surrounded by a passive shield made of
plexiglas ($6-13$ cm), high purity copper (thickness $3-6$ cm) and
lead (15 cm).

The event-by-event data acquisition records information on the
amplitude (energy) and arrival time of each detector event. The
energy resolution of the detector (FWHM) for $\gamma$ lines with
the energies 60 keV ($^{241}$Am), 570 keV and 1064 keV
($^{207}$Bi) was measured as 37\%, 15\% and 10\%, respectively.
These data can be fitted by the function:
FWHM$_{\gamma}=-8.5+\sqrt{16\cdot E_{\gamma}}$, where
FWHM$_{\gamma}$ and the energy of $\gamma$ quanta $E_{\gamma}$
are in keV.

The energy spectrum measured with the ZnWO$_{4}$ crystal during
429 h in the low background set-up is presented in Fig.~5. The
spectra of CaWO$_4$ and CdWO$_4$ scintillators measured in similar
conditions are given for comparison (the spectra are normalized by
their measurement times and detector masses). The
background of the ZnWO$_4$ detector is substantially lower than
that of the CaWO$_4$ and is comparable with that of the CdWO$_4$
above $\approx0.5$ MeV. Note, that below 0.5 MeV the counting rate
of the ZnWO$_4$ detector is one order of magnitude lower than that of
CdWO$_4$. Obviously, it is due to presence in the CdWO$_4$
crystals of the $\beta$ active $^{113}$Cd isotope (natural
abundance of $\approx12\%$).

\nopagebreak
\begin{figure}[ht]
\begin{center}
\mbox{\epsfig{figure=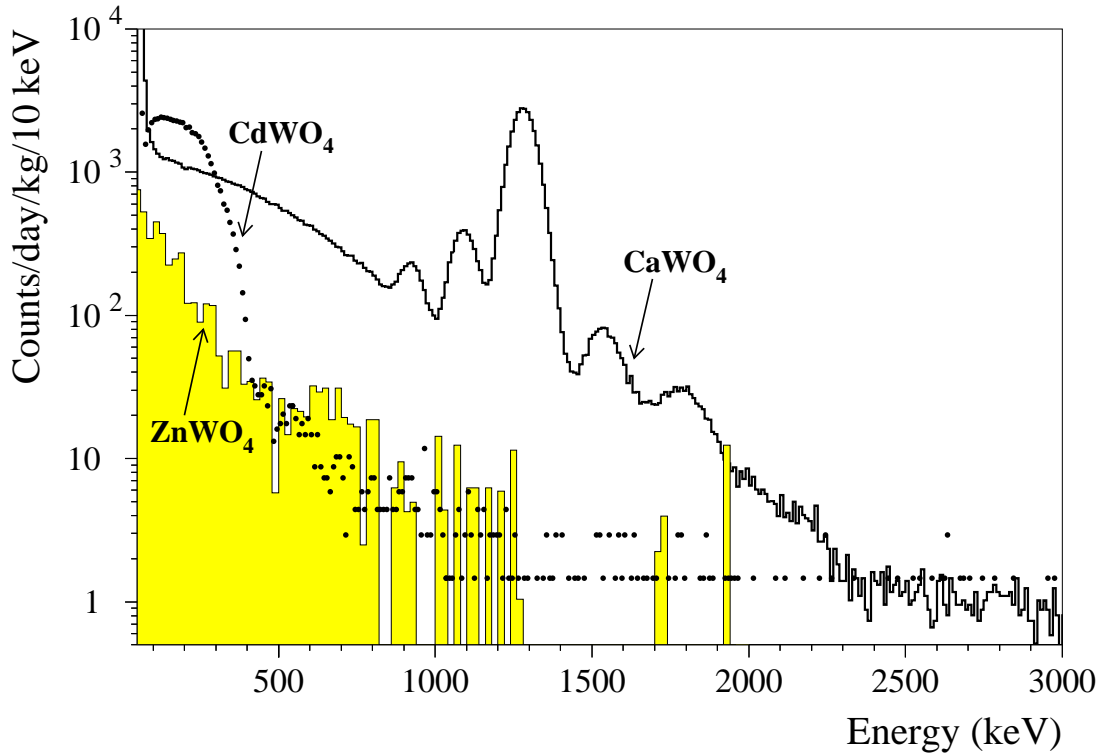,height=10.0cm}}
\caption {Energy spectra of ZnWO$_4$ (4.5 g, 429 h),
CaWO$_{4}$ (189 g, 1734 h), and CdWO$_4$ (448 g, 37 h)
scintillation crystals measured in the low background set-up. The
CaWO$_4$ crystal is considerably polluted by radionuclides from
U-Th chains (see for details \cite{Zde04}). Beta decay of
$^{113}$Cd ($Q_{\beta}=316$ keV, $T_{1/2}=7.7\times 10^{15}$ yr)
dominates in the low energy part of the CdWO$_4$ spectrum. The
background of the ZnWO$_4$ detector is caused mainly by external
$\gamma$ rays.}
\end{center}
\end{figure}

The spectrum of the ZnWO$_4$ detector was simulated with the
GEANT4 package \cite{GEANT}.
Initial kinematics of particles emitted in $\alpha$ and $\beta$
decays of nuclei was generated with the DECAY event generator
\cite{DECAY}.
The background is caused mainly by
$\gamma$ quanta from the PMT. In the spectrum there are no
peculiarities which could be referred to the internal trace
radioactivity. Therefore, only limits on contaminations of this
crystal by nuclides from U--Th families as well as by $^{40}$K,
$^{90}$Sr--$^{90}$Y, $^{137}$Cs, and $^{147}$Sm were set on the
basis of the experimental data. With this aim the spectrum was
fitted in different energy intervals by simple model composed of
an exponential function (to describe external $\gamma$ rays) and
background components searched for (simulated with the GEANT4
package). Because equilibrium of U--Th families in crystals is
expected to be broken, different part of the families were
considered separately. Results of these estimations are presented
in Table 3.

\begin{table}[htb]
\caption{Radioactive contaminations in ZnWO$_4$, CaWO$_4$, and
CdWO$_4$ crystal scintillators.}
\begin{center}
\begin{tabular}{|l|l|l|l|l|}
\hline
 Chain           & Source                      & \multicolumn{3}{|c|}{Activity (mBq/kg)}     \\
 \cline{3-5}
 ~               &                             & ZnWO$_4$   &CaWO$_4$ \cite{Zde04}  & CdWO$_4$  \cite{Geo96,Bur96,Dan96,Dan03}  \\
 \hline
 ~ & ~ & ~ & ~ & ~ \\
 $^{232}$Th      & $^{232}$Th                  & $\leq 3.3$ & 0.69(10)        & 0.053(5)     \\
 ~               & $^{228}$Th                  & $\leq 0.2$ & 0.6(2)          & $\leq 0.004-0.039(2)$ \\
 ~ & ~ & ~ & ~ & ~ \\
 $^{235}$U       & $^{227}$Ac                  & $\leq 0.2$& 1.6(3)          & 0.0014(9)    \\
 ~ & ~ & ~ & ~ & ~ \\
 $^{238}$U       & $^{238}$U                   & $\leq 3.2$   &14.0(5)      & $\leq 0.6$   \\
 ~               & $^{230}$Th                  & $\leq 4.5$  &--            & $\leq 0.5$   \\
 ~               & $^{226}$Ra                  &  $\leq 0.4$    & 5.6(5)      & $\leq 0.004$  \\
 ~               & $^{210}$Pb                  & $\leq 1$ &  $\leq 430$       & $\leq 0.4$   \\
 ~               & $^{210}$Po                  &  ~   &  291(5)    & ~ \\
 ~ & ~ & ~ & ~ & ~ \\
 ~               & $^{40}$K                    &   $\leq 12$ &$\leq 12$     & 0.3(1)       \\
 ~               & $^{90}$Sr                   &  $\leq 1.2$ & $\leq 70$      & $\leq 0.2$   \\
 ~               & $^{113}$Cd                  & --         &  --             & 580(20)      \\
 ~               & $^{113m}$Cd                 & --         &  --             & 1-30         \\
 ~               & $^{137}$Cs                  & $\leq 20$ &  $\leq 0.8$      & $\leq 0.3-0.43(6)$ \\
 ~               & $^{147}$Sm                  &  $\leq 1.8$   &  0.49(4)     & $\leq 0.04$  \\
 ~ & ~ & ~ & ~ & ~ \\
 \hline
\end{tabular}
\end{center}
\end{table}

More sensitive limits on radioactive impurities associated with
the daughters of $^{232}$Th, $^{235}$U and $^{238}$U were obtained
with the help of the time-amplitude analysis described in detail
in \cite{Dan95,GSO}. For example, the fast sequence of two
$\alpha$ decays from the $^{232}$Th family was searched for:
$^{220}$Rn ($Q_\alpha =6.41$ MeV, ${\it T}_{1/2}=55.6$ s)
$\rightarrow $ $^{216}$Po ($Q_\alpha =6.91$ MeV, $T_{1/2}=0.145$
s) $\rightarrow $ $^{212}$Pb (which are in equilibrium with
$^{228}$Th). No events were selected by using this method and the
limit $\leq 0.2$ mBq/kg on the activity of $^{228}$Th was set. In
the same way, the absence of the fast sequences $^{214}$Bi
($Q_{\beta}=3.27$ MeV, $T_{1/2}=19.9$ m) $\rightarrow $ $^{214}$Po
($Q_{\alpha}=7.83$~MeV, $T_{1/2}=164~\mu$s) $\rightarrow $
$^{210}$Pb from $^{238}$U family, and $^{219}$Rn ($Q_\alpha =6.95$
MeV, $T_{1/2}=3.96$ s) $\rightarrow $ $^{215}$Po ($Q_\alpha =7.53$
MeV, $T_{1/2}=1.78$ ms)$\rightarrow $ $^{211}$Pb  from $^{235}$U
chain leads to the following limits: $^{226}$Ra $\leq 0.4$ mBq/kg,
and $^{227}$Ac $\leq 0.2$ mBq/kg.

\section{DISCUSSION}

\subsection{Search for 2$\beta$ processes in zinc and tungsten}

Double $\beta$ decay of two zinc ($^{64}$Zn and $^{70}$Zn)
and two tungsten ($^{180}$W and $^{186}$W) isotopes can be studied
with the help of ZnWO$_{4}$ detector. Their mass differences,
isotopic abundances, possible decay channels are listed in Table
4. We performed preliminary investigation to check a possibility
of such experiments using the data of the low background
measurements with the ZnWO$_4$ scintillator $\oslash 14\times 4$
mm.

\begin{table}[htb]
\caption{Half-life limits on 2$\beta$ processes in  zinc and
tungsten isotopes. Mass difference \cite{Aud95} and isotopic
abundance \cite{abundance} are denoted as $\Delta M$ and
$\delta$, respectively.}
\begin{center}
\begin{tabular}{|c|l|l|l|l|}
\hline
 Transition        & Decay     & Decay & \multicolumn{2}{c|}{Experimental $T_{1/2}$ limit, yr} \\
 \cline{4-5}
 $\Delta M$ in keV & channel   & mode  & Present work    & Previous limits \\
 $\delta$ in \%    & ~         & ~     &  90\% (68\%) CL & ~ \\
 \hline
 $^{64}$Zn$ \rightarrow ^{64}$Ni & $2\varepsilon$        & $0\nu$   & $0.7~(1.0)\times10^{18}$    & $8\times10^{15}$ \cite{Ber53} \\
 1096.4(0.9)                     & $\varepsilon \beta^+$ & $2\nu$   & $4.3~(8.9)\times10^{18}$    & $2.3\times10^{18}$ (68\% CL) \cite{Nor85} \\
 48.63(0.60)                     & $\varepsilon \beta^+$ & $0\nu$   & $2.4~(3.6)\times10^{18}$    & $2.3\times10^{18}$ (68\% CL) \cite{Nor85} \\
 ~  &  ~  & ~ & ~ & $T_{1/2}=(1.1\pm0.9)\times10^{19}$ yr \cite{Bik95} \\
 ~  &  ~ & ~ & ~ & ~ \\
 $^{70}$Zn$ \rightarrow ^{70}$Ge & $2\beta^-$           & $2\nu$    & $1.3~(2.1)\times10^{16}$    & -- \\
 1000.9(3.4)                     & $2\beta^-$           & $0\nu$    & $0.7~(1.4)\times10^{18}$    & $1.3\times10^{16}$ (90\% CL) \cite{Kiel03} \\
 0.62(0.03)  &  ~   & ~ &   ~ & ~ \\
   ~  &  ~ & ~ & ~ & ~ \\
 $^{180}$W$ \rightarrow ^{180}$Hf   & $2K$              & $2\nu$    & $0.7~(0.8)\times10^{16}$   & $0.7~(0.8)\times10^{17}$ 90\%~(68\%) CL \cite{Dan03} \\
 146(5)                             & $2\varepsilon$    & $0\nu$    & $0.9~(1.1)\times10^{16}$  & $0.9~(1.3)\times10^{17}$ 90\%~(68\%) CL \cite{Dan03} \\
 0.12(0.01)  &  ~   & ~ &   ~ & ~ \\
  ~              &  ~              & ~ & ~ & ~ \\
 $^{186}$W$ \rightarrow ^{186}$Os  &  $2\beta^-$            & $2\nu$   & $1.4~(2.5)\times10^{18}$   & $3.7~(5.3)\times10^{18}$ 90\%~(68\%) CL \cite{Dan03} \\
 488.0(1.7)           & $2\beta^-$ & $0\nu$   & $1.1~(1.7)\times10^{19}$   & $1.1~(2.1)\times10^{21}$ 90\%~(68\%) CL \cite{Dan03} \\
 28.43(0.19)     &  ~   & ~ &   ~ & ~ \\
\hline
\end{tabular}
\end{center}
\end{table}

In the energy spectrum accumulated with the ZnWO$_4$ crystal (see
Fig.~6) there are no peculiarities which can be attributed to
$2\beta$ processes in the mentioned zinc or tungsten
isotopes.\footnote{The response functions of the ZnWO$_4$ detector
expected for the different channels and modes of 2$\beta$
processes in zinc and tungsten were simulated with the
help of the DECAY \cite{DECAY} and
GEANT4 \cite{GEANT} codes.} Therefore only lower
half-life limits can be set according to formula: lim$T_{1/2}=N
\cdot \eta \cdot t \cdot $~ln$2/$lim$S$, where $N$ is the number
of potentially $2\beta$ unstable nuclei, $\eta $ is the detection
efficiency, $t$ is the measuring time and lim$S$ is the number of
events of the effect searched for which can be excluded with a
given confidence level (CL).

\nopagebreak
\begin{figure}[ht]
\begin{center}
\mbox{\epsfig{figure=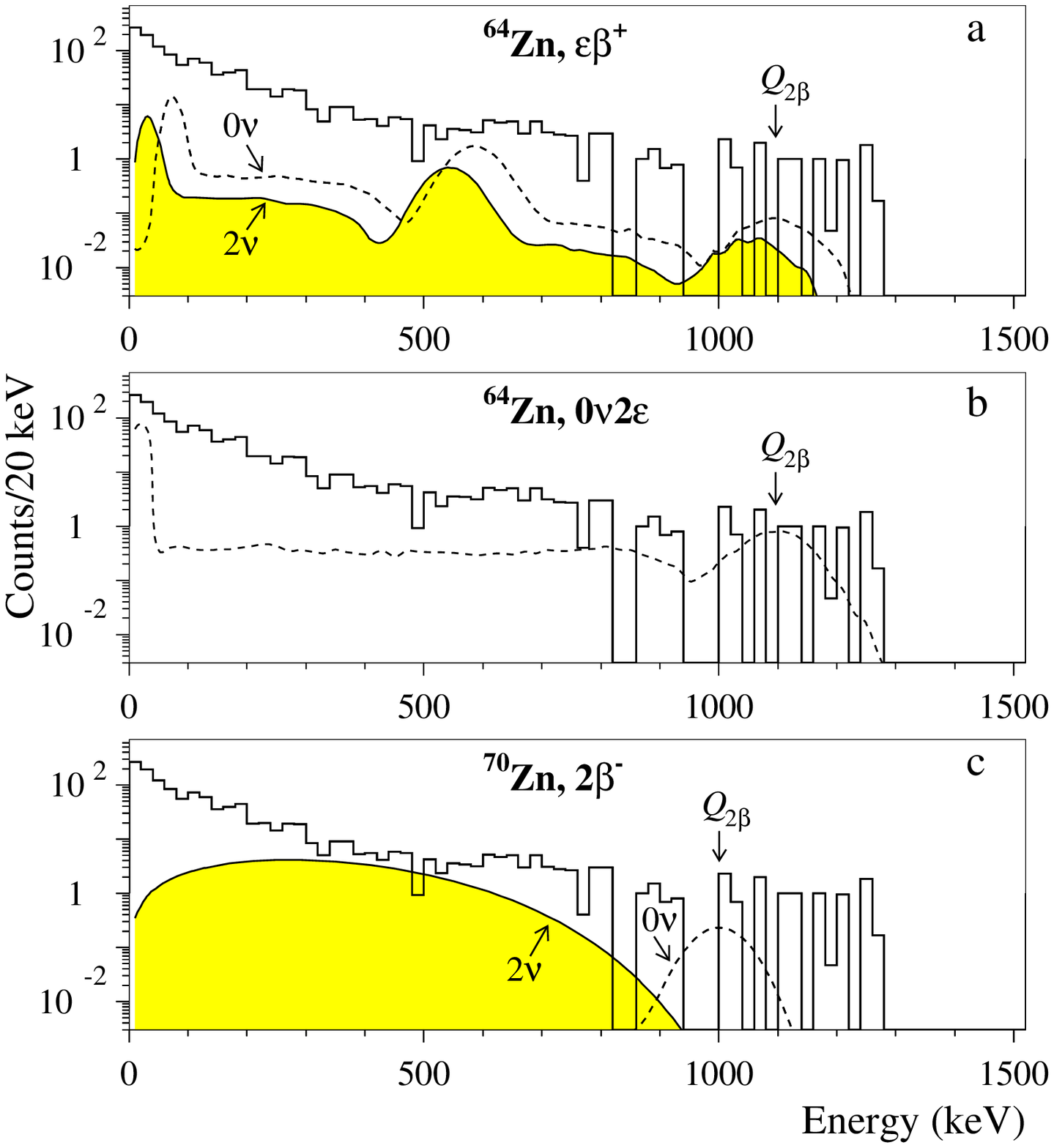,height=10.0cm}}
\caption {The background energy spectrum of ZnWO$_4$
scintillation crystal (4.5 g of mass, 429 h measurements in the
Solotvina Underground Laboratory) together with the
excluded at 68\% CL distributions for $\varepsilon\beta^+$ (a) and
$0\nu2\varepsilon$ (b) processes in $^{64}$Zn. (c) The excluded at
68\% CL distributions for the $2\beta^-$ decays of $^{70}$Zn.}
\end{center}
\end{figure}

{\bf $^{64}$Zn ($\varepsilon\beta^+$)}. The energy of the first
excited level of $^{64}$Ni (1346 keV) is higher than the $Q_{2\beta}$
value of $^{64}$Zn, therefore transitions only to the ground state
of $^{64}$Ni are allowed. To estimate the value of lim$S$ for the
two neutrino mode of electron capture in $^{64}$Zn with positron
emission ($2\nu\varepsilon \beta^+$), the experimental spectrum
was fitted in the energy interval $380-800$ keV. The simple model
including the effect searched for and an exponential function (to
describe the background) was chosen. The least squares fit
($\chi^2/n.d.f.=21.5/19=1.1$) gives total area of the effect
$-15\pm29$ counts which corresponds (in accordance with the
Particle Data Group recommendations \cite{PDG}) to lim$S=34~(16)$
counts at 90\%~(68\%) CL. Taking into account $\approx100$\%
registration efficiency and number of $^{64}$Zn nuclei in the
crystal ($N=4.21\times{10}^{21}$), one can calculate the half-life
limit:

\begin{center}
 $T_{1/2}^{2\nu\varepsilon \beta^+} \geq {4.2~(8.9){\times}10^{18}}$ yr at 90\%~(68\%) CL.
\end{center}
In the same way the half-life bound on the neutrinoless mode was
set:

\begin{center}
 $T_{1/2}^{0\nu\varepsilon \beta^+} \geq {2.4~(3.6){\times}10^{18}}$ yr at 90\%~(68\%) CL.
\end{center}

The experimental spectrum of the ZnWO$_4$ crystal in the energy
interval $50-1500$ keV is presented in Fig.~6(a) together with
excluded at 68\% CL distributions of the $2\nu$ and $0\nu$
$\varepsilon \beta^+$ decay of $^{64}$Zn. Despite the very little
detector mass and short exposure used in the present work, the
obtained limits are higher than those reported before
($T_{1/2}\geq2.3\times10^{18}$ yr at 68\% CL for $2\nu$ and
$0\nu$ processes \cite{Nor85}).

Recently, an experimental observation of the
$\varepsilon\beta^{+}$ decay of $^{64}$Zn with
$T_{1/2}^{(0\nu+2\nu)\varepsilon\beta^+}=(1.1\pm 0.9)\times10^{19}$ yr
was claimed in
\cite{Bik95}. A $\oslash7.6\times 7.6$ cm NaI(Tl) scintillation
and a 25\% efficiency HPGe detector, operated in coincidence, were
used in the experiment. The excess of $\approx85$ events in the
511 keV peak was observed with zinc sample (mass of 350 g, 392 h
of exposure), while no effect was detected without sample or with
copper (iron) blanks. Unfortunately, possible radioactive
contamination of the zinc sample (which could give the observed
excess of events) and of the Cu and Fe blanks were not discussed
in \cite{Bik95}, and, to our knowledge,
no further experiments were carried out to
prove or elaborate this result.

We have estimated the sensitivity of an experiment with about one
kg ZnWO$_4$ crystal\footnote{Even larger zinc tungstate crystals
($\oslash60\times150$ mm, that is of $\approx3$ kg) can be
produced \cite{Grab84}.} ($\approx10^{24}$ of $^{64}$Zn
nuclei) to $\varepsilon\beta^{+}$ decay of $^{64}$Zn.
As it was shown by the GEANT4 simulation,
the registration efficiency in the high energy part of the
expected $\varepsilon \beta^+$ distributions will be substantially
increased (at least by an order of magnitude). Supposing
additional decrease of background by a factor of $\approx5$ (due
to larger detector, improvement of the set-up, and
time-amplitude and pulse shape analyses), we expect to reach
after $\approx1$ yr of exposure the sensitivity (in terms of the
half-life limit) of $\approx 5\times10^{21}$ yr both for the $0\nu$ and
$2\nu$ modes. Thus, the $2\nu\varepsilon \beta^+$ decay of
$^{64}$Zn at the level of $10^{20}$ yr could be surely observed,
proving or disproving the result of \cite{Bik95}. It should be
noted, that the $0\nu$ and $2\nu$ modes of the $\varepsilon\beta^+$
process in $^{64}$Zn can be distinguished with a ZnWO$_4$
detector, while it is impossible to do in experiments similar to
\cite{Bik95}, where external zinc sample was used.

{\bf $^{64}$Zn ($2\varepsilon$)}. For the $2\nu $ double electron
capture in $^{64}$Zn from $K$ shell the total energy released in a
detector is equal to $2E_K=16.7$ keV (where $E_K=8.3$ keV is the
binding energy of electrons on $K$-shell of nickel atoms).
Detection of such a little energy deposit requires rather low
energy threshold. In our measurements the energy threshold for
acquisition was not enough low to search for $2\nu2K$ decay in
$^{64}$Zn. However, as one can see on Fig.~1(b), X-rays with the
energy of a few keV can be resolved from PMT noise.

At the same time, the neutrinoless mode of $2\varepsilon$ process
in $^{64}$Zn can be searched for using the existing data. In such
a process all available energy has to be transferred to gamma
quanta, conversion electrons, X-rays or Auger electrons.
Corresponding simulated distribution is shown in Fig.~6(b)
where we supposed transfer of energy to one gamma quantum,
the most disadvantageous process with the point of view of
efficiency of detection.
The maximum likelihood fit in the energy interval
400--1300 keV gives $59\pm86$ counts for the effect searched for,
which leads to the half-life limit:

\begin{center}
 $T_{1/2}^{0\nu2K} \geq {0.7~(1.0){\times}10^{17}}$ yr at 90\%~(68\%) CL.
\end{center}
This limit is higher than that obtained in  \cite{Ber53}.

It should be stressed, that $^{64}$Zn is one of only few
potentially $2\varepsilon$/$\varepsilon\beta^+$ unstable nuclides
with rather high natural abundance (see, for instance, review
\cite{DBD-table}). Therefore large scale experiment to search for
double beta processes in $^{64}$Zn can be realized without using
of high cost enriched isotopes.

{\bf $^{70}$Zn} (2$\beta^-$). The peak at 1001 keV with energy resolution
FWHM$~=118$ keV is expected in the spectrum of the
ZnWO$_4$ detector for the $0\nu2\beta$ decay of $^{70}$Zn. Number
of $^{70}$Zn nuclei in the crystal is 5.4$\times{10}^{19}$,
registration efficiency $\eta\approx100\%$. Maximum likelihood fit
of the background spectra in the energy range $0.7-1.3$ MeV gives
$-0.6\pm1.9$ counts for the area of the effect searched for, which
corresponds (again using the recommendations \cite{PDG}) to
lim$S=2.6~(1.3)$ counts at 90\%~(68\%) CL. It allows us to
restrict the half-life of $^{70}$Zn relatively to the $0\nu2\beta$
decay:

\begin{center}
 $T_{1/2}^{0\nu2\beta} \geq {0.9~(1.2)\times 10^{18}}$ yr at 90\%~(68\%) CL.
\end{center}
Similarly, the lower half-life limit on the $2\nu2\beta$ decay of
$^{70}$Zn decay was established as:
\begin{center}
 $T_{1/2}^{2\nu2\beta} \geq {1.3~(2.1)\times 10^{16}}$ yr at 90\%~(68\%) CL.
\end{center}
The excluded at 68\% CL distributions corresponding to $2\beta$ decays of
$^{70}$Zn are presented in Fig.~6(c).

Despite modest level of these results, the limit on the
neutrinoless mode of $2\beta$ decay of $^{70}$Zn is higher than
that obtained in recent experiment \cite{Kiel03} while the
$2\nu$ mode is elaborated for the first time.

{\bf $^{180}$W and $^{186}$W.} We have also estimated the
half-life limits on the $2\beta$ decay of $^{180}$W and $^{186}$W
on the basis of the measurements with the ZnWO$_4$ crystal. All
the limits (see Table 4) are lower than those set earlier with 330
g CdWO$_4$ detector operated during much longer time
(692 h for $^{180}$W and 13316 h for $^{186}$W)
\cite{Dan02,Dan03}. However, the
advantage of the ZnWO$_4$ detector (as compared with CdWO$_4$) is
absence of the $\beta$ active $^{113}$Cd and $2\nu2\beta$ active
$^{116}$Cd isotopes. As a result, background of the ZnWO$_4$
detector in the energy interval of interest is much lower.
Therefore sensitivity to search for the double beta processes in
$^{180}$W and $^{186}$W could be improved with a larger ZnWO$_4$
detector.

\subsection{$\alpha$ decay of $^{180}$W}

It should be also mentioned a potentiality of ZnWO$_4$ detector
(scintillator or bolometer) to measure alpha decay of $^{180}$W,
indication of which (with the half-life $\approx 10^{18}$ yr)
has been obtained in the low background experiment with CdWO$_4$
crystal scintillators \cite{W-alpha} and confirmed recently in the
measurements with CaWO$_4$ as scintillator \cite{Zde04} and
bolometer \cite{Coz04}.

\subsection{ZnWO$_4$ as detector for the dark matter particles search}

The CaWO$_4$ crystals are promising material for a dark matter
detector. Powerful background suppression was realized in
\cite{CRESST,ROSEBUD} due to simultaneous registration of heat and
light signals. However, radioactive contaminations of the crystal
used in \cite{CRESST,ROSEBUD}, as well as investigated in work
\cite{Zde04}, are rather high. Radioactive contamination of the
ZnWO$_4$ crystal is essentially lower than that of CaWO$_4$.
Although there are no data about performance of ZnWO$_4$ crystals
as low temperature particle detectors, ZnWO$_4$ is expected to be
an applicable material for bolometric technique taking into
account the encouraging results obtained with physically analogous
CdWO$_4$ \cite{Ale94} and CaWO$_4$ \cite{CRESST,ROSEBUD} crystals.

If so, ZnWO$_4$ crystals could provide an interesting possibility
to search for spin-dependent inelastic scattering of WIMP with
excitation of low energy nuclear levels (see Table 5 where
nuclides with nonzero spin presented in ZnWO$_4$ crystals are
listed). Identification of such "mixed" (nuclear recoil plus
$\gamma$ quanta) events could be possible due to the simultaneous
registration of heat and light signals. A heat/light ratio for
such events would differ from "pure" nuclear recoils or
$\gamma$($\beta$) events. For example, search for spin-dependent
inelastic scattering of WIMP with excitation of the $^{67}$Zn
and/or $^{183}$W low-energy nuclear levels (93 keV and 47 keV,
respectively) can be realized with ZnWO$_4$ crystals. Detection of
nuclear recoils and delayed (with the half-life 9~$\mu$s in the
case of $^{67}$Zn) $\gamma$ quanta could give a strong signature
for the process searched for.

\begin{table}[htb]
\caption{Nuclides with nonzero spin present in ZnWO$_4$ crystals
(data from Ref. \cite{Fir96})}
\begin{center}
\begin{tabular}{|c|c|l|l|l|}
\hline
 Nuclide &  Spin, parity & Isotopic abundance (\%)  &  \multicolumn{2}{c|}{First excited level}\\
 \cline{4-5} ~    &  ~   &  ~   & energy (keV) & life time \\
 \hline
 $^{17}$O   & $5/2^+$ & 0.038  & 870.7 & 179 ps \\
 $^{67}$Zn  & $5/2^-$ & 4.1    & 93.3 & 9.16 $\mu$s  \\
 $^{183}$W  & $1/2^-$ & 14.3    & 46.5 & 0.188 ns \\
\hline
\end{tabular}
\end{center}
\end{table}

In addition, the observed with ZnWO$_4$ scintillators dependence
of pulse shape on the direction of $\alpha$ particles irradiation
relative to the crystal axes (supposing that such a dependence
would remain valid for nuclear recoils) could be used to detect a
diurnal asymmetry of WIMP direction \cite{DM}.

\section{CONCLUSIONS}

Scintillation properties of ZnWO$_4$ crystals were studied.  The
energy resolution 9.1\% (662 keV $^{137}$Cs $\gamma$ line) was
obtained with ZnWO$_4$ crystal scintillator placed in liquid and
viewed by two PMTs. Shapes of scintillation signals was
investigated, and clear pulse-shape discrimination for
$\gamma$($\beta$) and $\alpha$ events was achieved. Dependences of
the $\alpha/\beta$ ratio and scintillation pulse shape on the
energy of alpha particles were measured. It was also found that the
$\alpha/\beta$ ratio and pulse shape show the dependence on
direction of $\alpha$ irradiation relatively to the main crystal
planes of ZnWO$_4$ crystal.

Radioactive contamination of ZnWO$_4$ crystal was measured in the
Solotvina Underground Laboratory. No radioactive pollution in
ZnWO$_4$ scintillator was observed. Limits on radioactive
contamination in ZnWO$_4$ crystal are on the level from tenth to a
few mBk/kg which is one--two order of magnitude lower than those
for CaWO$_4$.

Two potentially $2\beta$ active zinc and two tungsten isotopes
($^{64}$Zn, $^{70}$Zn, $^{180}$W, and $^{186}$W) were studied with
the help of ZnWO$_4$ crystal. New improved half-life limits on
double beta decay in zinc isotopes were established. In
particular, for the $2\nu \varepsilon\beta^+$ decay of $^{64}$Zn
the half-life limit was set as $T_{1/2}^{2\nu
\varepsilon\beta^+}\geq 8.9\times 10^{18}$ yr at 68\% CL.

Due to a low level of radioactive contamination, ZnWO$_4$ crystals
seems to be encouraging material for the dark matter experiments.
ZnWO$_4$ crystals could provide a possibility to search for
spin-dependent inelastic scattering of WIMP with excitation of low
energy nuclear levels. Most interesting targets are $^{67}$Zn and
$^{183}$W.

A strong signature can be used to search for spin-dependent
inelastic scattering of WIMP on nuclei with nonzero spin (which
are present in ZnWO$_4$ crystals), via registration of nuclear
recoils and delayed $\gamma$ quanta. Due to dependence of ZnWO$_4$
scintillation pulse shape on direction of irradiation by high
ionizing particles, there is $in~principle$ a possibility to
search for diurnal modulation of WIMP direction. It could be a
strong signature for WIMP evidence.


\end{document}